%Paper: hep-ph/9503254
%From: Kari Juhani Eskola <kjeskola@pcu.helsinki.fi>
%Date: Tue, 7 Mar 1995 14:28:36 +0200
%Date (revised): Thu, 9 Mar 1995 12:54:27 +0200
%Date (revised): Thu, 9 Mar 1995 13:31:12 +0200

%%%%%%%%%%%%%%%%%%%%%%%%%%%%%%%%%%%%%%%%%%%%%%%%%%%%%%%%%%%%%%%%
% HU-TFT-95-17
% Formation and Evolution of Quark-Gluon PLasma at RHIC and LHC
%%%%%%%%%%%%%%%%%%%%%%%%%%%%%%%%%%%%%%%%%%%%%%%%%%%%%%%%%%%%%%%%
\def\lsim{\,\raise0.3ex\hbox{$<$\kern-0.75em\raise-1.1ex\hbox{$\sim$}}\,}
\def\gsim{\,\raise0.3ex\hbox{$>$\kern-0.75em\raise-1.1ex\hbox{$\sim$}}\,}

\def\gev{\,{\rm GeV}}
\def\fm{\,{\rm fm}}
\font\smallrm=cmr8

\magnification=1200
\hsize=16.0truecm
\vsize=23.0truecm

\pageno=0
{\nopagenumbers
\line{\hfill Preprint HU-TFT-95-17}
\line{\hfill hep-ph/9503254}
\line{\hfill 7 March 1995}

\vskip 2.5cm

\centerline{\bf  FORMATION AND EVOLUTION OF QUARK-GLUON PLASMA}
\centerline{\bf  AT RHIC AND LHC$^*$}
\vskip 1cm
\centerline{\bf K.J. Eskola}
\centerline{Laboratory of High Energy Physics,}
\centerline{P.O. Box 9, 00014 University of Helsinki, Finland}
\centerline{email: kjeskola@ fltxa.helsinki.fi}
\vskip 1.5cm
\centerline{\bf Abstract}
\vskip 0.6cm
Initial conditions for quark-gluon plasma formation at $\tau=0.1\fm$
are considered in lowest order perturbative QCD. Chemical
composition, thermalization of the formed semihard quark-gluon system
and especially implications of the new HERA
parton distributions with the enhancement at small $x$ are studied.
The plasma at $\tau=0.1\fm$ is shown to be strongly gluon dominated
both at RHIC and LHC, and a possibility for rapid thermalization of
gluons at LHC is pointed out.  Uncertainties in the
calculations, particularly shadowing corrections to the
parton distributions, are discussed. Free streaming and ideal hydro limits
for the evolution of the gluonic plasma with the calculated minijet
initial conditions are demonstrated,
and a lower limit for final multiplicities
obtained for the LHC nuclear collisions.
\vfill
\hrule\medskip
{\smallrm\baselineskip 10pt
\noindent
$^*$ Plenary talk at {\it Quark Matter '95},
Eleventh Conference on Ultra-Relativistic Nucleus-Nucleus Collisions,
Monterey, CA, U.S.A., January 9-13, 1995. }
\eject
}

\baselineskip=13pt
%%%\pageno=1
{}~~\vskip 2 truecm
\noindent
\line{{\bf FORMATION AND EVOLUTION OF QUARK-GLUON PLASMA}\hfil\break}
\line{{\bf AT RHIC AND LHC}\hfil}
\bigskip\bigskip\noindent
K. J. Eskola
\bigskip\noindent
Laboratory of High Energy Physics, P.O. Box 9,
FIN-00014 University of Helsinki, Finland
\bigskip\bigskip
Initial conditions for quark-gluon plasma formation at $\tau=0.1\fm$
are considered in lowest order perturbative QCD. Chemical
composition, thermalization of the formed semihard quark-gluon system
and especially implications of the new HERA
parton distributions with the enhancement at small $x$ are studied.
The plasma at $\tau=0.1\fm$ is shown to be strongly gluon dominated
both at RHIC and LHC, and a possibility for rapid thermalization of
gluons at LHC is pointed out.  Uncertainties in the
calculations, particularly shadowing corrections to the
parton distributions, are discussed. Free streaming and ideal hydro limits
for the evolution of the gluonic plasma with the calculated minijet
initial conditions are demonstrated,
and a lower limit for final multiplicities
obtained for the LHC nuclear collisions.
\bigskip\medskip
\noindent{\bf 1.\ INTRODUCTION}
\medskip
During  recent years, a lot of effort has been devoted to studying
the space-time evolution of ultrarelativistic heavy ion collisions.
Especially, from the point of view of quark-gluon plasma (QGP) formation,
it is important to understand
energy deposition and particle production mechanims in the central
rapidity region of the collisions.
Qualitatively these mechanisms can be divided into two categories:
semihard and soft.

Semihard particle production is based on perturbative QCD (pQCD).
The semihard gluons, quarks and antiquarks are produced
in parton-parton collisions and they have transverse momenta
$p_T\ge p_0\sim 1...3$ GeV. These semihard QCD quanta are often called
minijets [1] because even in $pp(\bar p)$ collisions
they are not observable as individual jets  below $p_T \sim 5\gev$ [2].
However, even if minijets are not at all directly observable in
nuclear collisions, they are expected to play a major role in the
space-time evolution of central collisions at collider energies
[3-7]
Below the momentum transfer scale $\sim 1-2$ GeV
pQCD  is not expected to be reliable anymore,
and modeling for soft particle production
from beam jet fragmentation is needed. In heavy ion collisions
this can be done e.g. in terms of
strings as in hadron-hadron collisions [8], or in terms of
a strong background  color field  decaying into particles through the
Schwinger mechanism [9].

In the space-time evolution of an ultrarelativistic heavy ion collision
the semihard and soft particle
production mechanisms take place at different time scales.
Semihard QCD quanta in the central rapidity region
are  typically formed at very early times, at
$\tau\sim 1/p_T\lsim 1/p_0\sim 0.1$ fm, whereas the
characteristic time scale for the soft processes (like the decay time
of the background color field), is $\tau\sim 1$ fm. In this sense, the
minijet system is formed first and it will then serve as initial
condition for the further evolution of the system. An example of
this kind of modeling can be found e.g. in ref. [10].

In the central rapidity region the relative strength of the semihard
and soft particle production mechanisms depends strongly on the
cms energy of the heavy ion
collision. The semihard part is expected to play a more important role
with increasing energy, so that it is negligible at energies
$\sqrt s\lsim 20\,A$GeV, becomes relevant at the Brookhaven
Relativistic Heavy Ion Collider (RHIC) energies
$\sqrt s\sim 200\,A$GeV, and finally dominates at the CERN
Large Hadron Collider (LHC) energies $\sqrt s\sim 5.5\,A$TeV
[3,4,6,11].
This conclusion is based on the idea of multiple minijet production
 as the cause of the rapidly rising inelastic and total cross sections of
$pp$ and $p\bar p$ collisions at energies $\sqrt s\gsim 200$ GeV.
A detailed discussion and further references can be found e.g. in ref [11].

Since the first ideas of the importance of the semihard processes
in ultrarelativistic heavy ion collisions [3,5], a lot of inspiring
work has been done on creating event generators like HIJING [6]
and Parton Cascade Model (PCM) [7]. Naturally, the techniques
of treating semihard processes in  event generators like PYTHIA [12]
for hadron-hadron collisions  has been of great help in building up
heavy ion event generators. In spite of the differences in initialization
and in final state interactions, the semihard processes lie in the
heart of generating events with e.g. HIJING and PCM at heavy ion collider
energies. Therefore, to have a better control over the results from
simulations, it is important to understand the basic processes
as well as possible, and especially to study the uncertainties, like
nuclear and hadronic shadowing of the parton distributions,
higher order contributions to the cross sections, and the cut off
scale $p_0$ used to  give the division between semihard and soft physics.

In this talk, I will concentrate on the semihard parton production only.
I will first estimate the initial conditions for early QGP formation.
The main emphasis  will be in consequences of the small-$x$ enhancement of
the parton distributions observed in the HERA measurements,
as studied in ref [13].
After considering the uncertainties in the pQCD calculation,
I will discuss the evolution of the formed QGP with minijet
initial conditions, and finally I will give an estimate for the lower limit
of multiplicities for LHC nuclear collisions.

\vfill
\eject
\noindent
{\bf 2.\ FORMATION OF QGP}
\medskip
\noindent
{\bf 2.1.\ Impact of the HERA measurements on minijet and $E_T$ production in
$AA$ collisions}
\medskip

The new deep inelastic $ep$-data from HERA [14] exhibit an increase
in the structure function $F_2^p(x,Q^2)$ at small $x$
relative to the behavior implied by earlier data. Distribution
function analyses using the new data have been carried out.
These analyses, in particular, constrain the gluon distribution function,
which is not directly measured. The estimates to be presented in this talk
are computed with parton distributions D0', D-' and H by Martin, Roberts
and Stirling (MRS) [15,16]. The
corresponding gluon distributions for these sets at $Q=2$ GeV are shown in
Fig. 1. At small $x$ the distributions shown behave as $x^{-\delta}$ with
$\delta = 0$, $-0.3$ and $-0.5$, for D0', H and D-', correspondingly.
In the global analysis by MRS, the set H follows from the best fit to
the HERA data [16].

\midinsert
\vskip 8.0 truecm
%\hskip 4.0 truecm

\includegraphics{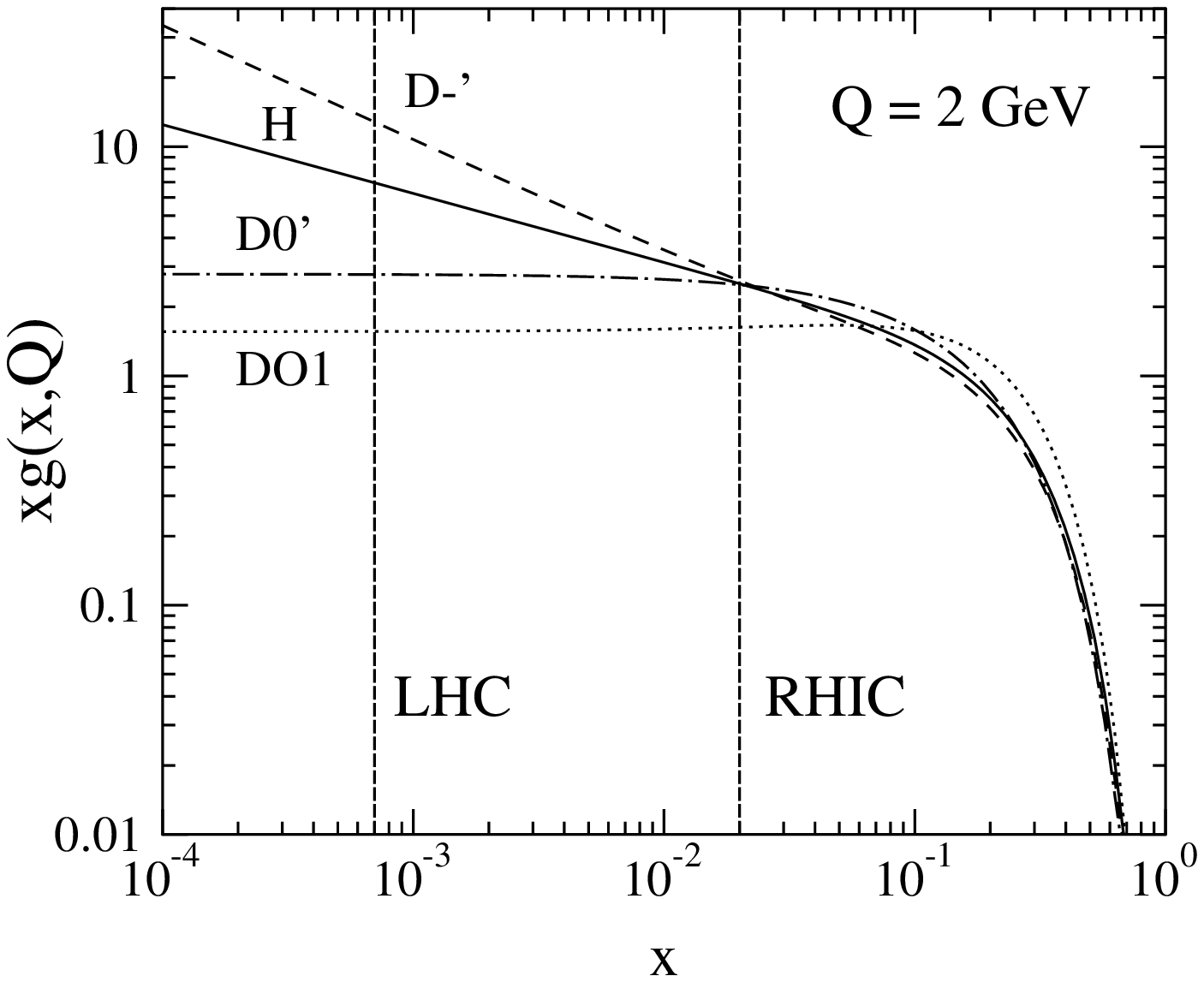}

\smallskip
\baselineskip=10.0pt plus 1.0pt minus 0.5pt  % 12 +0 -0
{\noindent {\bf Fig.~1.}
Gluon distribution functions MRS D0', D-, and H [15,16]
at the scale $Q=2$ GeV. The old gluon distribution of
Duke-Owens set 1 (DO1) [17] is shown for comparison.
For the vertical lines, see the text.}
\endinsert
\smallskip

For minijet production at high energies the consequences of the
observed small-$x$ rise are obvious. Minijets falling into the
mid-rapidity $y=0$ dominantly have transverse momenta
$p_T\sim p_0\sim$ 2 GeV, and they mainly come from momentum
fractions $x\sim 2p_0/\sqrt s$. In Fig. 1, the vertical lines illustrate
the typical value of $x$ for LHC and RHIC with $\sqrt s=200\gev$ and
$\sqrt s=5500\gev$, correspondingly. Difference between the
MRS distributions shown is relatively small at RHIC, whereas at LHC there
is a sizable difference between the sets, and this is
reflected into the minijet cross sections discussed below.

When considering the number of produced semihard gluons, quarks and antiquarks,
both in hadronic and nuclear collisions, the key quantity is the
integrated (mini)jet cross section
$$
\sigma_{\rm jet}(\sqrt s,p_0)=\int_{p_0^2}^{s/4}dp_T^2dy_1dy_2 {1\over2}
{d\sigma_{\rm jet} \over dp_T^2dy_1dy_2}, \eqno(1)
$$
where the inclusive differential jet  cross section is given by
$$
{d\sigma_{\rm jet} \over dp_T^2dy_1dy_2} =
K\sum_{{ijkl=}\atop{q,\bar q,g}}
x_1f_{i/p}(x_1,p_T^2) x_2f_{j/p}(x_2,p_T^2)
{d\hat\sigma\over d\hat t}^{ij\rightarrow kl}.\eqno(2)
$$
The momentum transfer in the process $ij\rightarrow kl$ is $p_T$,
the fractional momenta of the partons $i$ and $j$ are $x_{1,2}$, and
$y_{1,2}$ are the rapities of the partons $k$ and $l$.
The calculation is only done to leading order  (${\cal O}(\alpha_s^2$)).
The higher order terms are in principle available
from the calculation of S.D. Ellis-Kunszt-Soper (EKS) [18] but here
these are simulated simply by  a constant factor $K\sim 2$,
which seems to be an acceptable approximation with a  scale choice
$Q=p_T$ [19].

In the eikonal formulation
(see e.g. [11] and references therein) we can interpret
$2\sigma_{\rm jet}/\sigma_{\rm in}^{pp}$ as the
average number of semihard partons with $p_T\ge p_0$ produced in
an inelastic $pp$ collision. This ratio may become much larger than 2,
(see e.g. $\sigma_{\rm jet}$ shown in fig. 1 in [13])
especially at  LHC energies with small-$x$ enhanced gluon distributions,
which indicates that one is effectively including minijet pairs
not only from
multiple (independent) semihard collisions but also contributions from the
initial partonic flux, i.e. from the gluon ladders leading to the
virtuality $Q\sim p_T$ [20]. However, in the central rapidity,
the pairs should dominantly come from the semihard collisions.

In the first approximation, where all nuclear modifications to the
parton densities $f_i$ are neglected, the average number of
minijets produced in an $AA$ collision with an  impact parameter ${\bf b}$
is obtained from
$$
\bar N_{AA}({\bf b},\sqrt s,p_0) =
T_{AA}({\bf b})\, 2\sigma_{\rm jet}(\sqrt s,p_0)\eqno(3)
$$
where $T_{AA}({\bf b})$ is the standard nuclear overlap function.
Here we will need only
$T_{AA}({\bf 0})\approx A^2/(\pi R_A^2)\approx 32/{\rm mb}$ for $A = Pb$.
Notice that there are two reasons for large $\bar N_{AA}$ for heavy nuclei:
the overlap function scales as $\sim A^{4/3}$ and the perturbative
cross section $\sigma_{\rm jet}$ becomes large at high energies.

The results for $\sigma_{\rm jet}$ integrated over all rapidities can be
found in [13], but what is of greater interest for us here, are the
numbers of minijets falling into the central rapidity unit $|y|\le 0.5$.
By making this cut in Eq. (1) (as  will be done in Eq. (4))
and by careful bookkeeping
of the flavors which fall into this rapidity window, we find out an
estimate for the chemical  composition of the produced minijets at
LHC and at RHIC. This is shown in Table 1.
\smallskip
\midinsert
\moveright2.5cm\vbox{
\halign{
#\hfil&\hfil#\qquad&\hfil#\qquad&\hfil#\qquad&\hfil#\qquad\cr
\noalign{\vskip6pt \hrule width8.65truecm \vskip6pt}
Set&total&$g$&$q$&$\bar q$\cr
\noalign{\vskip6pt \hrule width8.65truecm \vskip6pt}
DO1\quad&\quad 776&\quad640&\quad72&\quad64\cr
{}\quad&\quad 134&\quad100&\quad21&\quad13\cr
{}&{}&{}&{}&{}\cr
D0'\quad&\quad 1510&\quad1250&\quad136&\quad128\cr
{}\quad&\quad 207&\quad162&\quad27&\quad18\cr
{}&{}&{}&{}&{}\cr
H\quad&\quad 3250&\quad2710&\quad276&\quad266\cr
{}\quad&\quad 200&\quad157&\quad26&\quad17\cr
{}&{}&{}&{}&{}\cr
D-'\quad&\quad5980&\quad5220&\quad385&\quad373\cr
{}\quad&\quad200&\quad157&\quad26&\quad17\cr
\noalign{\vskip6pt \hrule width8.65truecm \vskip6pt}}}
\smallskip
\baselineskip=10.0pt plus 1.0pt minus 0.5pt  % 12 +0 -0
{
\noindent {\bf Table.~1.}
Values of $\bar N_{AA}({\bf 0},\sqrt s,p_0)$ calculated for
$p_0=2\gev$, $\sqrt s=200\gev$, $5500\gev$ and for the
three parton distributions considered in the text.
The upper numbers are for LHC and the lower ones for RHIC.
The results with Duke-Owens set 1 distributions (DO1)
are shown for comparison.
An overall factor $K=2$ but no shadowing is included. Both the total as well
as the contribution from gluons, quarks and antiquarks is given.
}
\endinsert
\smallskip

We make two observations:
First,
as expected, the small-$x$ rise of the parton distributions
does not affect much the numbers for RHIC,
but inreases considerably the numbers for LHC.
Secondly,
the formed minijet plasma is obviously strongly gluon dominated:
about 80 \% of the produced partons are gluons. At this point,
one should bear in mind that the figures above are only the lowest order
estimates, and the number of partons is well defined. However,
when considering higher order terms in the jet cross section,
one encounters collinear singularities in the case of
three-jet final states, and, without a  definition
of a jet size $R$ in the $(y,\phi)$-plane
it is not clear when the partons should be counted as separate jets and when
as one jet [18]. In general,
radiative effects would tend to increase the gluon
dominance of the formed minijet plasma.

Let us then consider perturbatively the transverse energy
in the central rapidity region of $AA$  collisions. This can be estimated
(see [4] for details) in terms of the first $p_T$ moment
$$
 \sigma_{\rm jet}(\sqrt s, p_0)\langle E_T\rangle=\int_{p_0} dp_Tdy_1dy_2
{d\phi\over2\pi}
{1\over2}{d\sigma_{\rm jet} \over dp_Tdy_1dy_2}(\epsilon_1+\epsilon_2)p_T,
\eqno(4)
$$
where the rapidity cut is made by the functions $\epsilon_k$:
$\epsilon_k=\epsilon(p_T,y_k)=1$ if the parton $k$ falls inside the
region of our interest and =0 otherwise. To illustrate the $p_0$ dependence,
$\sigma_{\rm jet}\langle E_T\rangle $ is plotted as a function of $p_0$
in Fig. 2. Note again that no shadowing is included in the calculation.
The composition into contributions from gluons, quarks and antiquarks
with $p_0 = 2$ GeV is shown in Table 2. The estimate for the  total initial
$E_T$ in one unit of $y$ near $y=0$ is then
$$
\eqalign{
\bar E_T^{PbPb}({\bf 0})(\sqrt s,p_0,|y|\le0.5)
&=T_{PbPb}({\bf 0})\,\sigma_{\rm jet}(\sqrt s,p_0)\langle E_T\rangle\cr
&=\,\,\,\,3200\quad (387)\, {\rm GeV}\qquad {\rm  DO1:}\, xg(x)\sim{\rm
const},\cr
&=\,\,\,\,5440\quad (563)\, {\rm GeV}\qquad {\rm  D0'\,\,\,:}\, xg(x)\sim{\rm
const},\cr
&=\,10300\quad(547)\, {\rm GeV}\qquad {\rm H\,\,\,\,\,\,\,\,:}\, xg(x)\sim
x^{-0.3},\cr
&=\,17600\quad (538) \,{\rm GeV}\qquad  {\rm D-'\,\,:}\, xg(x) \sim x^{-0.5}.
\cr}\eqno(5)
$$
\noindent where the numbers are for LHC (RHIC), correspondingly.

\midinsert
\vskip 8.0 truecm
%\hskip 4.0 truecm

\includegraphics{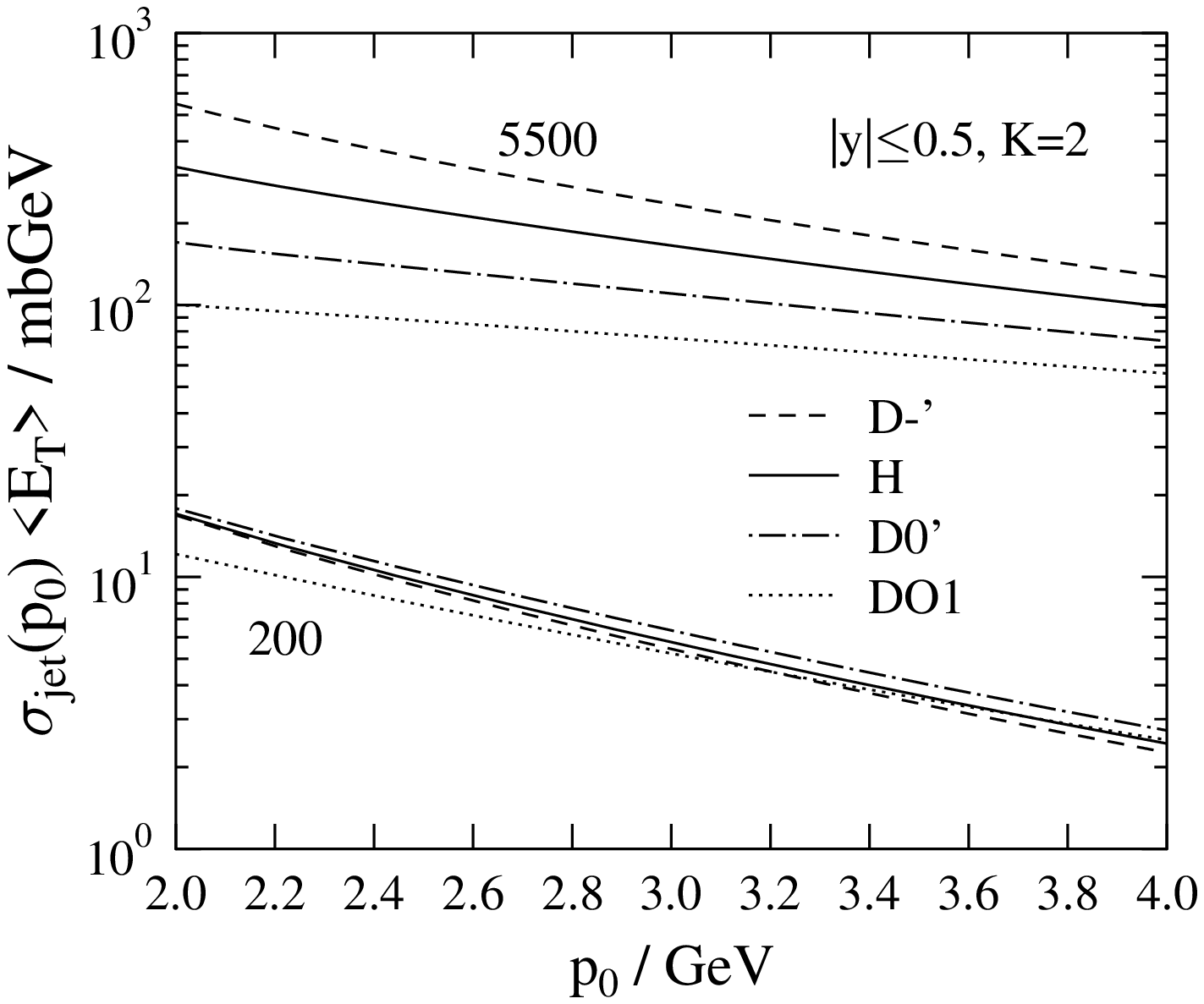}

\smallskip
\baselineskip=10.0pt plus 1.0pt minus 0.5pt  % 12 +0 -0
{\noindent {\bf Fig.~2.}
The first $p_T$ moment $\sigma_{\rm jet}(\sqrt s,p_0) \langle E_T \rangle$
from Eq.~(4) with the sets of distribution functions marked
on the figure and for $\sqrt{s}=200$ and 5500 GeV [13].
}
\endinsert
\smallskip

Again,  the increase in $E_T$ associated with the small-$x$ increase is
significant for LHC. However, this $E_T$ is the final $E_T$ only if no further
interactions take place in the system. The largest decrease in $E_T$ is
obtained if the system thermalizes and an ideal hydro flow is generated.
Another interesting point is that in spite of the apparently large values
of $E_T$ in Eq. (5), at most only about 1.5\% of the total
energy is stopped perturbatively.

\smallskip
\midinsert
\moveright2.5cm\vbox{
\halign{
#\hfil&\hfil#\qquad&\hfil#\qquad&\hfil#\qquad&\hfil#\qquad\cr
\noalign{\vskip6pt \hrule width9.0truecm \vskip6pt}
Set&$\sigma_{\rm jet}\langle E_T\rangle$&$g$&$q$&$\bar q$\cr
\noalign{\vskip6pt \hrule width9.0truecm \vskip6pt}
DO1\quad&\quad 100&\quad84.5&\quad8.29&\quad7.41\cr
{}\quad&\quad 12.1&\quad9.16&\quad1.87&\quad1.11\cr
{}&{}&{}&{}&{}\cr
D0'\quad&\quad 170&\quad142&\quad14.5&\quad13.6\cr
{}\quad&\quad 17.6&\quad14.0&\quad2.12&\quad1.50\cr
{}&{}&{}&{}&{}\cr
H\quad&\quad 322&\quad270&\quad26.7&\quad25.5\cr
{}\quad&\quad 17.1&\quad13.3&\quad2.30&\quad1.47\cr
{}&{}&{}&{}&{}\cr
D-'\quad&\quad549&\quad479&\quad35.8&\quad34.5\cr
{}\quad&\quad16.8&\quad13.2&\quad2.24&\quad1.40\cr
\noalign{\vskip6pt \hrule width9.0truecm \vskip6pt}}}
\smallskip
\baselineskip=10.0pt plus 1.0pt minus 0.5pt  % 12 +0 -0
{
\noindent {\bf Table.~2.}
As Table 1, but for $\sigma_{\rm jet}(p_0)\langle E_T \rangle$
in units of mbGeV with
$|y|\le 0.5$, $\sqrt s = 200\gev,\, 5500\gev$ and $p_0=2\gev$.
}
\endinsert
\smallskip

Since the longitudinal size of the comoving volume is
$\Delta z=\tau \Delta y$ and $\tau_i\sim 1/p_0=0.1\fm $
the obtainable energy densities can be estimated
with the Bjorken formula [21] as
$\epsilon(\tau_i)=\langle E_T^A(|y|<0.5)\rangle p_0/\pi R_A^2$.
Already without the small-$x$ enhancement of the parton distributions,
i.e. for distributions like  MRSD0', the energy densities
$\epsilon(\tau_i)$ become large as compared to
the critical density $\epsilon_c\sim 1...3$ GeVfm$^{-3}$ associated
with the deconfinement phase transition. For MRSD0' we get
$\epsilon(\tau_i)\sim$ 410 (43) GeVfm$^{-3}$ for LHC (RHIC).
An interesting observation
is that for distributions like MRSD0' at high enough energies
the initial energy density does {\it not}
depend strongly of $p_0$. This is because of the scalings:
$\sigma_{\rm jet}(p_0)\sim p_0^{-2}$, $\sigma_{\rm jet}(p_0)\langle E_T
\rangle\sim p_0^{-1}$, and the $p_0$-dependence in $\epsilon(\tau_i)$
is canceled out.
With the small-$x$ enhancement of the parton distributions, the results
depend more on $p_0$, as illustrated in Fig.~2.

About the approximations done, one should note that the transit times
of the colliding nuclei were neglected. In $AA$ collisions with
$A\sim 200$ at RHIC the transit time is about 0.2 fm,
which will reduce the  estimate for the maximum energy density
by a factor $\sim$ 2 [10]. For LHC, the
transit time is negligible as compared to $1/p_0=0.1$ fm, but nuclear
shadowing may reduce the estimate at least by a similar factor
(see e.g. [34]).
\bigskip

\noindent
{\bf 2.2.\ Chemical equilibrium at $\tau = 0.1$ fm?}
\medskip
The lowest order pQCD results of [13] described in the previous
chapter show that the minijet-gluons dominate the early QGP. Let
us now study how far from chemical equilibrium we actually
are initially at $\tau\sim$ 0.1 fm.

For an ideal gas of massless gluons, quarks and antiquarks,
we have the following ratios at zero chemical potential:
$$
{n_g\over n_q+n_{\bar q}}\bigg|_{\rm ideal} = {16\over 9N_f}\le
0.9,\,\,\,\,\,\,\,\
{\epsilon_g\over \epsilon_q+\epsilon_{\bar q}}\bigg|_{\rm ideal} =
{32\over 21N_f}\le 0.8,
\,\,\,\,\,\,\,N_f\ge 2\eqno(6)
$$
{}From the previous lowest order pQCD results, we can extract the ratios as
$$
{n_g\over n_q+n_{\bar q}}\bigg|_{\rm pQCD}
\sim 3.7\,({\rm RHIC}) ..... 5({\rm LHC}), \,\,\,\,\,\,\,\,
{\epsilon_g\over \epsilon_q+\epsilon_{\bar q}}\bigg|_{\rm pQCD}
\sim 3.5\,({\rm RHIC})....5({\rm LHC})\eqno(7)
$$
The conclusion is the same from both number and energy densities
and for both RHIC and LHC:
the pQCD ratios  are far from the equilibrium ones.
Many more quarks and antiquarks would be needed for complete
chemical equilibrium.

Then, an interesting question is that if we consider gluons only,
can they be in chemical equilibrium, i.e. thermalized
at  $\tau\sim 0.1$ fm? In this case
it would mean that the gluon number density is
close to the thermal density obtained by assuming that all the initial energy
density, $\epsilon_g(\tau_i)$, is fully thermalized.
Therefore, for LHC, we compare  energy/gluon computed in two ways:

\noindent
i) take the previous pQCD results from tables 1 and 2 to get
  $\bar E_{T,g}^{PbPb}$ and $\bar N_{PbPb}^g$
with $p_0=2 \gev$, $|y|\le 0.5$.

\noindent
ii) take the pQCD estimate for $\bar E_{T,g}^{PbPb}$,
convert this into equilibrium  ($\mu_g=0$) energy density
$\epsilon_g(\tau_i)= 3a{T_i}^4$, where $a=16\pi^2/90$,
 solve for $T_i$, and get energy/gluon as
$\epsilon_g(\tau_i)/n_g(\tau_i)\approx 2.7 T_i$.
For LHC, the results for different parton distributions are
$$
\eqalign{
E/{\rm gluon}
&=4.1\,{\rm GeV}\qquad 2.7T_i=2.0\,\gev\quad{\rm DO1,}\cr
&=3.6\,\gev\qquad\qquad    \, =2.2\,\gev\quad {\rm D0',}\cr
&=3.2\,\gev\qquad\qquad    \, =2.6\,\gev\quad {\rm H,}\cr
&=2.9\,\gev\qquad\qquad    \, =3.0\,\gev\quad {\rm D-'}.\cr }
\eqno(8)
$$
For the distributions with $xg(x)\sim{\rm const}$ the thermalization of
gluons requires a degradation of their average energy by collisions.
For RHIC this is  also the case [22].
With the HERA parton distributions the secondary interactions are only
needed for changing the directions of momenta to make the distribution
uniform. The conclusion is that  the assumption of fast initial
thermalization seems reasonable at LHC when new HERA
parton densities are used.

\bigskip
\noindent
{\bf 2.3.\ Uncertainties: $K$, $p_0$ and shadowing}
\medskip
In the previous results for the jet cross sections there are
three main uncertainties: the $K$-factor, the momentum cut-off $p_0$
and shadowing of the parton distributions.

The calculations were done only to the lowest order (LO) pQCD, and a
factor $K=2$ was used together with  a scale choice $Q=p_T$ to
simulate the contributions from higher orders of $\alpha_s$.
Here one should note that a '$K$-factor' always depends on
the scale choice; it can even be removed by choosing the scale differently.
The Born level jet cross section does {\it not} depend on the jet size
$R$ in the $(y,\phi)$-plane,
while the next-to-leading order (NLO) calculation [18] {\it does}, and
the experimentally measured jet cross sections {\it do}.
Therefore, the $K$-factor
always depends on the jet size, regardless of whether $K$ is defined by
comparing the LO results to the inclusive jet data [23] or
to the LO+NLO calculation. Naturally, $K$ also depends on the parton
distributions and may also depend on the cms energy.

Although I am not discussing the inclusive jet production here but, rather,
considering more semi-inclusive quantities, (i.e.  all minijets and their
transverse energy in the central rapidity unit), I show
in Fig. 3 the inclusive 1-jet distribution from [19]
in $pp$ collisions at $\sqrt s=200\gev$ with $Q=p_T$ and $R=1$.
This shows that with this scale choice, it is a good approximation to have
constant $K\sim2$ to simulate the NLO terms in the jet cross sections.

\midinsert
\vskip 9.0 truecm
%\hskip 4.0 truecm

\includegraphics{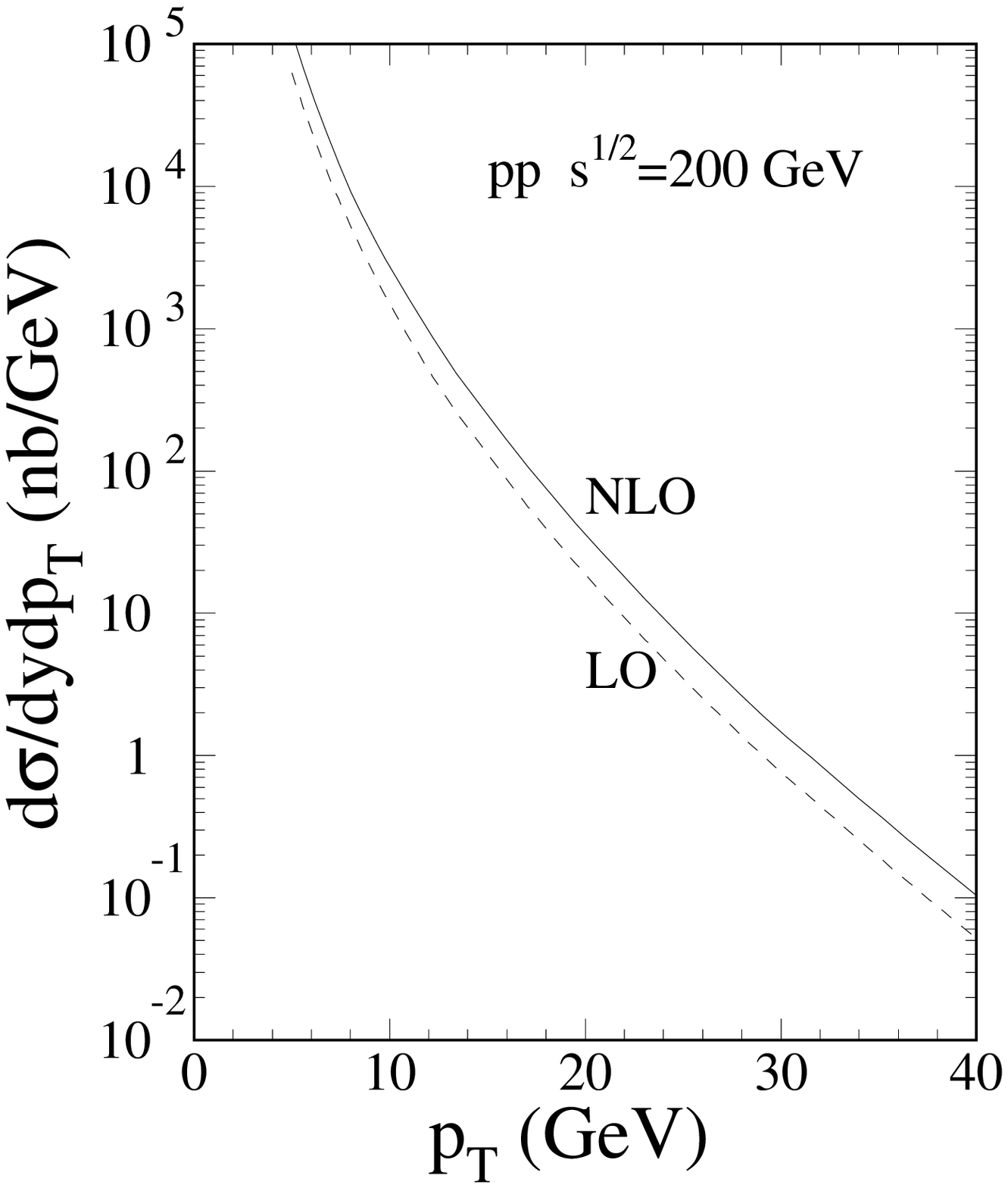}

\smallskip
\baselineskip=10.0pt plus 1.0pt minus 0.5pt  % 12 +0 -0
{
\noindent {\bf Fig.~3.}
The inclusive one-jet cross section $d\sigma/dp_Tdy$ vs.
$p_T$ at $y=0$ in $pp$ collisions at $\sqrt s=200\gev$,
as predicted by the NLO  (solid) and LO(dashed) calculations
with $Q=p_T$. The MRSD-' parton distributions were used.
For the NLO calculation $R=1.0$. The figure is from
ref. [19], and the calculation was done by using
the program by EKS [18].
}
\endinsert
\smallskip

When the NLO contributions to the jet cross sections are
included, the dependence on the scale choice should be weaker.
This has been nicely demonstrated in [18] (see also [19]),
and it works well for jets with $p_T\gsim 10 \gev$.
However, when approaching the minijet region
$p_T\sim p_0\sim 2\gev$, the convergence of the perturbation series becomes
worse, and one would eventually need even higher order terms in the jet
cross sections in order to improve the determination of the scale choice.
This does not mean that the LO minijet results would be
incorrect, it simply means that it may be very difficult to pin down the
accuracy of the minijet calculation below a factor 2. As an example of this,
we see in Fig. 3 that at least down to  $p_T=5\gev$
the jet NLO calculation does not exhibit any striking behavior
as compared to the LO result, even though at this low $p_T$ the
NLO result depends on the scale choice practically as much as the LO does.
However,  it is clear that this problem needs further studies.

In the approach considered here, the parameter $p_0$ defines
a division into semihard and soft physics. Therefore,
$p_0$ is a phenomenological parameter, and data on multiplicities and
cross sections are needed to determine $p_0$.
In the eikonal approach for the $p\bar p (p)$ collisions,
the cross section $\sigma_{\rm soft}$ for
processes with $p_T < p_0$ is modeled in together with
$\sigma_{\rm jet}(p_0)$. Through the condition that experimentally
measured $p\bar p (p)$ total cross sections are reproduced,
the value of $\sigma_{\rm jet}(p_0)$ then depends on
the choice for $\sigma_{\rm soft}$ [1,11].
In principle, in determination
of $p_0$, one should also compare the final multiplicities as was done in
[11]. Rather than doing a complete analysis like this,
I have shown the  results from [13]
as a function  of $p_0$. However, a reanalysis on determination of
$p_0$ with the new  HERA parton distributions should be done,
and the possible $A$ dependence of $p_0$ should be studied in more detail.

Related to the cut-off $p_0$, I note the following.
Part of the higher order contributions come from
$2\rightarrow 2+(n\ge1)$ processes. At the tree level, these have been
studied recently by E. Shuryak and Li Xiong [24], based on the
matrix elements in [25]. In this treatment, one
is faced with the jet resolution problem mentioned before, and with cut-off
$s_{ij}>s_0\sim 4\gev^2$ for each pair $ij$, the higher order tree
graphs seem to become dominant over the $2\rightarrow2$ processes
at LHC energies [24]. Again, this shows that the actual number of gluons
depends on the resolution, and is connected with color
screening as well. Certainly, this interesting problem needs further
studies, especially for LHC energies.

Let us then come to the uncertainty due to modifications of the parton
distributions. Since here we are mostly interested in the small-$x$ phenomena,
I will only consider shadowing.

{\bf Shadowing in a proton} means a depletion or a saturation
in the gluon and sea quark distributions at $x\lsim 0.01$.
For the distributions  behaving as $xg\sim x^{-\delta}$, at sufficiently
small values of $x$ and/or $Q$  the partons start to overlap spatially,
and a reduction in the parton density results due to the occurring
recombinations.  Corrections from the gluon fusions to the
Dokshitzer-Gribov-Lipatov-Altarelli-Parisi (DGLAP) scale evolution [26]
at small $x$ have been calculated in the leading double logarithm approximation
by Gribov-Levin-Ryskin and Mueller-Qiu (GLRMQ)
[27]. In the semiclassical approximation it has been shown that
after correcting the evolution of the steeply rising gluon distributions
$\sim x^{-\delta}$, they  behave at asymptotically small
$x$ as $xg\sim\, {const}$ [28].
At RHIC energies, the shadowing corrections in proton can obviously
be neglected  but for LHC energies they should be taken into account,
at least with the new HERA parton  distributions.

There are two sources of gluon recombinations causing the
shadowing in a proton: the two fusing
gluon ladders, which couple four gluons to two gluons, can arise either from
independent constituents of a proton or from the same one, as discussed
in [27-29]. The HERA data for the structure function $F_2^p$
would seem to favor the first possibility; the latter source may
be ruled out as too strong a fusion since no saturation in $F_2$ is seen
so far. However, since the sea quarks are emitted from the gluons they carry
smaller $x$ than the mother gluons, and the saturation of the quark
distributions should show up at somewhat smaller $x$ than that of gluons.
It would be important to understand this better, since nuclear gluon
shadowing arises perturbatively from the same sources as in the proton
shadowing, with possibly different relative strengths [30].

Nuclear effects in the parton distributions in general
have been widely studied during the recent years (see e.g. [31]
for further references). However, a unified description through the
whole $x$-range is very difficult to make, due to the
 different origin of the effects at different ranges of $x$.
{}From the point of view of phenomena discussed  in this talk,
shadowing as the small-$x$ effect is the most important one.
Notice also that an  interesting theoretical idea for finding
out the nuclear gluon densities at small $x$
was discussed in this conference by R. Venugopalan [32].

{\bf Shadowing in a nucleus} means a depletion of the
nuclear parton densities relative to the parton densities in a proton:
$f_i^A(x,Q^2) <  Af_i^p(x,Q^2)$ at $x\lsim 0.1$. Unlike
shadowing in proton, nuclear shadowing of the structure function $F_2^A$
is experimentally clearly observed in deep inelastic scattering [33],
and it seems to be fairly independent of the scale $Q$.
Although measured in fixed target experiments, nuclear shadowing
gets a simple qualitative explanation in the infinite momentum frame in terms
of overlapping partons. If the longitudinal wavelength of a parton,
$\sim 1/(xp)$,  exceeds the contracted size of the nucleons
(or the inter-nucleon distance)  in the nucleus, $\sim 2r_n m_n/p$,
shadowing should show up at $x\lsim 1/(2m_nr_n)\sim 0.1$. According to the data
this seems to be  the case. Also, at
sufficiently small $x$ the partons from all nucleons at the same
transverse location will overlap, and a saturation of $f_i^A/Af_i^p$
is expected, at least when  $f_i^p$ starts saturating (see e.g. [30]).

Now, in the minijet calculation with the latest HERA parton distributions,
inclusion of shadowing is not a trivial matter. The nuclear
shadowing effects are often simulated by simply multiplying
the {\it unshadowed} parton distribution  in a proton by a
$Q$-independent ratio,  extracted from the data [33] for
$F_2^A/AF_2^p$.  While this is presumably a rather good approximation
for the sea quarks, it may not at all
be the right thing to do with the gluons.
Even when neglecting the shadowing in a proton -
which should be a reasonable approximation at the RHIC energies -
the ratio $xg_A/Axg_p$ may be more strongly $Q$-dependent
than that of the sea quarks. This was demonstrated in [34].
Besides the $Q$-dependence, one should also make sure that
the valence quark and momentum sum rules are fulfilled for the nuclear
parton distributions as well.

At LHC energies the problem is more complicated since the
gluon shadowing in a proton  must not be neglected.
Especially, with the rapidly rising gluon distributions, the
gluon density in a nucleus at small $x$
cannot be reliably approximated by simply multiplying by
$F_2^A/AF_2^p$ since $xg_p$ should be shadowed as well.
With the absence of a direct measurement for
the gluon densities,  the nuclear gluon shadowing can at the moment
only be modeled in
theoretically. By using the same perturbative mechanism to generate gluon
fusions as described in the proton case above, nuclear gluon shadowing
was studied in [30]. These results have not been implemented yet
into the minijet calculation, but the procedure is clear:
After extracting the nuclear parton distributions at some scale $Q$ as in [30]
and as in [34], one should then perform the  full DGLAP-GLRMQ
evolution in the case of nucleus. This work is in progress
[35].
To get a first feeling of the order of magnitude of the
shadowing effects, I refer to figs 5 and 7 of [34].

\vfill
\eject
\noindent
{\bf 3.\ EVOLUTION OF THE QGP}
\medskip
Let us then briefly consider the early evolution
of the QGP in a one-dimensional
longitudinal scaling expansion with boost invariant initial conditions [21].
The initial conditions are given by the system of formed semihard
gluons, quarks and antiquarks at $\tau\sim 0.1\fm$. As mentioned in the
Introduction and as shown in [10],
the soft processes are expected to be important
for the evolution of the energy density $\epsilon(\tau)$
up to RHIC energies but to be negligible at LHC energies
(see e.g. fig. 2 of [10]).
Since in this talk I am only considering the semihard contribution,
let me now comment on the evolution of the QGP at LHC energies only.

The QGP at $\tau=0.1\fm$ is strongly gluon dominated.
Let us therefore simplify the discussion by neglecting the fermions.
Furthermore, as demonstrated before, with the small-$x$ enhanced gluons
it is reasonable to approximate the gluons to be fully thermalized
at $\tau=0.1\fm$. Then, the evolution of the energy density $\epsilon_g(\tau)$
of the gluon plasma is governed by the equation [21]
$$
{d\epsilon_g\over d\tau} + {\epsilon_g+P_g\over\tau} = 0\eqno(9)
$$
where $P_g$ is the gluonic pressure.
If viscosity and Ohmic heating effects are taken into account,
they would appear on the right-hand side of the equation above.
These have been studied e.g.  in [10] in connection with
particle production from the  decaying
chromo-electric background field. However, for the
discussion below these effects are not relevant.

Now, there are two extreme cases for the evolution
of the gluon plasma:

\noindent i) free streaming:
			$P_g=0 \Rightarrow \epsilon_g\sim \tau^{-1}$

\noindent ii) ideal hydro:
			$P_g=\epsilon/3 \Rightarrow \epsilon_g\sim \tau^{-4/3}$

The assumption of full thermalization of glue is relevant only for the
latter case. If viscosity was included, it could at most reduce the $PdV$
work to zero, so the first case is clearly an upper bound for the evolution
of $\epsilon_g$ of
a purely gluonic plasma (GP). In Fig. 4 the evolution of  $\epsilon_g$ is
plotted as a function of $\tau$ in the two extreme cases mentioned above.
{}From the figure we see that if the deconfinement phase transition starts at
$\epsilon_g\sim 1-2{\gev}{\fm}^{-3}$, even the minimum life time of the
GP at LHC  is $\sim 10\fm$.

\midinsert
\vskip 8.0 truecm
%\hskip 4.0 truecm

\includegraphics{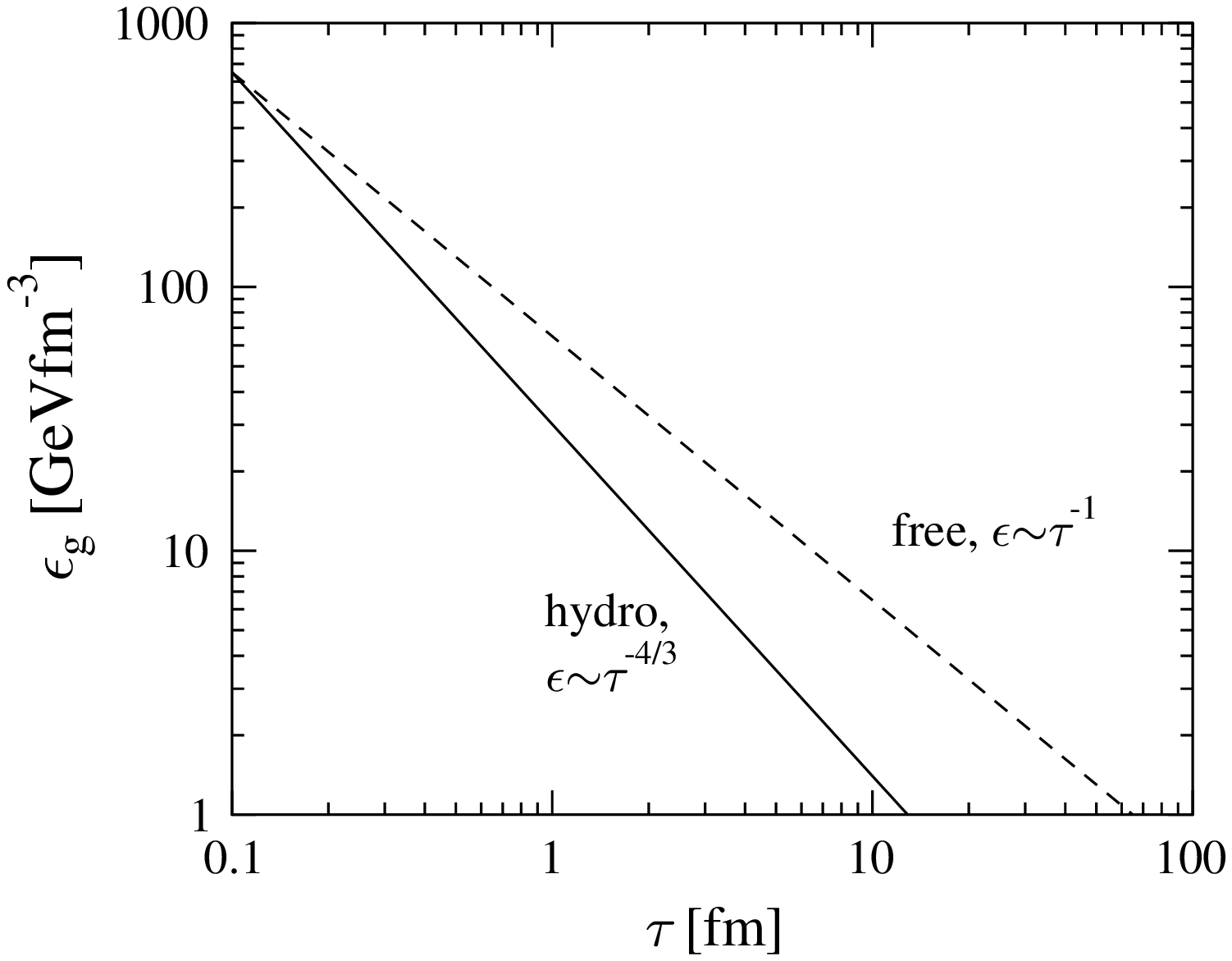}

\smallskip
\baselineskip=10.0pt plus 1.0pt minus 0.5pt  % 12 +0 -0
{\noindent {\bf Fig.~4.}
The two extreme cases of the evolution of the energy density $\epsilon_g$ of
the gluon plasma with initial  conditions given by the pQCD
calculation with the set MRSH [16].
Solid curve: ideal hydro flow with complete initial thermalization.
Dashed curve: free streaming of the initial gluon system.
}
\endinsert

\medskip
\noindent
{\bf 3.1.\ Multiplicities at LHC - lower limit}
\medskip
As a final topic, let me then estimate the multiplicities obtainable  at LHC.
The expected charged multiplicity is a crucial variable for the planning
of experiments [36] and the various theoretical predictions
lie roughly in the range 1500-8000.

Certainly, the free streaming case above would give us an
upper limit for the multiplicity, $\sim \bar E_T^{PbPb}(|y|\le0.5)/0.5\gev$.
This would correspond to having  no secondary interactions
after the initial
collision. In practice, this bound as such is of no realistic use,
since the final state interactions are practically inevitable
in this dense systems.
The lower limit of multiplicity is obtained by considering only the gluons
in the ideal hydro case (solid line in Fig. 4),
where the system is fully thermalized at
$\tau=0.1\fm$ and expands adiabatically thereafter. Then we only need to
compute the initial entropy and convert that into the final state multiplicity.
Pressure, energy density and entropy density are
$P_g=aT^4$, $\epsilon_g=3aT^4,$ $s_g=4aT^3$.
 Again, we estimate the longitudinal
size of the comoving volume as $\Delta z=\tau\Delta y$ and the
energy density at the central rapidity region becomes
$\epsilon_g(\tau_i)= \bar E_{T,g}^{Pb Pb}(|y|\le 0.5,p_0)p_0/(\pi R_{Pb}^2)$.
The charged final multiplicity is obtained as

$$
\eqalign{
{dN_{ch}\over dy}&\approx{2\over3}{1\over3.6}{dS\over dy}\approx
{2\over3}{1\over3.6}\pi R_A^2 4aT_i^3\tau_i\cr
&={2\over3}{4\over3.6}\biggl[{1\over27}\pi R_A^2a\tau_i
\{\bar E_{T,g}^{PbPb}(|y|<0.5)\}^3
\biggr]^{1/4} \cr
&=914\qquad  \,\,\,{\rm DO1,}\cr
&=1350\qquad  {\rm D0',}\cr
&=2180\qquad {\rm H, }\cr
&=3360\qquad {\rm D-'.}\cr
 }\eqno(10)
$$

It is interesting to notice how
crucial the lifetime $\tau_c$ of the (Q)GP-phase is
for the high multiplicities.
{}From Eq. (10)  it is observed that due to entropy conservation
$$
{dN_{ch}\over dy}\sim T_i^3\tau_i = T_c^3\tau_c\sim\tau_c,\eqno(11)
$$
since $T_c$ is in principle fixed by the theory. Therefore,
an increase in the initial parton production at $\tau_i=0.1\fm$
enhances the lower limit of the final multiplicity through
increasing the lifetime of the plasma.

\eject
\noindent
{\bf 4.\ CONCLUSIONS}
\medskip
I have discussed semihard transverse energy and parton production
in the mid-rapidity unit in central $PbPb$ collisions, and, considered
initial conditions for early QGP formation at $\tau=0.1\fm$ [13]. The main
emphasis of the talk has been on the consequences of the new HERA parton
distributions [15,16] to minijet production at the small scales $p_T\sim
2\gev$.
Especially, this lowest order pQCD calculation
shows that the early QGP is clearly gluon dominated both at RHIC and LHC.
For LHC, with the new, small-$x$ enhanced parton distributions
it was also shown that
rapid thermalization of the gluon system may be possible: initially
at $\tau=0.1\fm$ there are
enough gluons produced so that $\epsilon_g/n_g\sim 2.7 T_{\rm eq}$
as in an ideal gas of massless bosons. However,
the isotropization of the momenta what was not studied
in detail [22].

Uncertainties in the pQCD calculation of the initial
conditions were pointed out.
Especially corrections due to shadowing, both in a proton and
in a nucleus, were discussed,
but not yet implemented. For RHIC it seems that
the shadowing in a  proton may be neglected but
nuclear shadowing should be taken into account. For LHC,
since probing smaller $x$'s, one should include all shadowing effects
simultaneously [30]. Clearly, in this sense, RHIC will be
unaffected by the problems and uncertainties  in the parton distributions at
$x\lsim 0.01$, whereas LHC will be affected by these complications.
However, besides problems, the small-$x$ enhancement
causes also  larger initial energy and particle densities,
resulting in longer lifetime for the QGP and larger multiplicities.
A lower limit for multiplicites in the LHC nuclear collisions
was also estimated [13]. When the forthcoming colliders are in operation,
we will have a truly  interesting sequence of experiments,
AGS,SPS$ \rightarrow$ RHIC$ \rightarrow$ LHC to study, and
a wide range of multiplicities to understand, hopefully in terms of
a forming quark-gluon plasma.

\bigskip
\noindent{\bf Acknowledgements}

\noindent I would like to thank K. Kajantie, V. Ruuskanen, M. V\"anttinen and
 X.-N. Wang for  helpful discussions.

\medskip
\bigskip
\noindent
{\bf REFERENCES}
\medskip
\parindent=13pt

\item{[1]}
T. K. Gaisser and F. Halzen, Phys. Rev. Lett. 54 (1985)
1754;\hfil\break\noindent
L. Durand and H. Pi, Phys. Rev. Lett. 58 (1987) 303; \hfil\break\noindent
G. Pancheri and Y. N. Srivastava, Phys. Lett. B182 (1986) 199.

\item{[2]}
UA1 collaboration, C. Albajar {\it et al.}, Nucl. Phys. B309 (1988) 405.

\item{[3]}
K. Kajantie, P. V. Landshoff and J. Lindfors, Phys. Rev. Lett. 59 (1987) 2527.

\item{[4]}
K.J. Eskola, K. Kajantie and J. Lindfors, Nucl. Phys. B323 (1989) 37.

\item{[5]}
J.P. Blaizot, A.H. Mueller, Nucl. Phys. B289 (1987) 847.

\item{[6]}
X.-N. Wang and M. Gyulassy, Phys. Rev. D44 (1991) 3501;\hfil\break\noindent
Phys. Rev. D45 (1992) 844; Phys. Rev.  Lett. 68 (1992) 1480.

\item{[7]}
K. Geiger and B. M\"{u}ller, Nucl. Phys.  B369 (1992) 600;\hfil\break\noindent
K. Geiger, Phys. Rev. D47  (1993) 133.

\item{[8]}
G. Gustafson, Nucl. Phys. A566 (1994) 233,
in Proc. {\it Quark Matter '93},
eds. E. Stenlund, H.-{\AA}. Gustafsson, A. Oskarsson and I. Otterlund.

\item{[9]}
K. Kajantie and T. Matsui, Phys. Lett.  B164 (1985) 373;\hfil\break\noindent
G. Gatoff, A. K. Kerman and T. Matsui, Phys. Rev.  D36 (1987) 114.

\item{[10]}
K.J. Eskola and M. Gyulassy, Phys. Rev. C47 (1993) 2329.

\item{[11]}
X.-N. Wang, Phys. Rev. D43 (1991) 104.

\item{[12]}
T. Sj{\"o}strand and M. van Zijl, Phys. Rev. D36 (1987)
2019;\hfil\break\noindent
T. Sj{\"o}strand, Comp. Phys. Comm. 39 (1986) 347;\hfil\break\noindent
T. Sj{\"o}strand and M. Bengtsson, {\it ibid.} 43 (1987) 367.

\item{[13]}
K.J. Eskola, K. Kajantie and P. V. Ruuskanen, Phys. Lett. B332 (1994) 191.

\item{[14]}
H1 Collaboration, I. Abt {\it et al.}, Nucl. Phys. B407 (1993)
515;\hfil\break\noindent
ZEUS Collaboration, M. Derrick {\it et al.}, Phys. Lett. B316 (1993) 412.

\item{[15]}
A.D. Martin, W.J. Stirling and R.G. Roberts, Phys. Lett.
B306 (1993) 145.

\item{[16]}
A.D. Martin, W.J. Stirling and R.G. Roberts, RAL preprint
93-077 (1993).

\item{[17]}
D. W. Duke and J. F. Owens, Phys. Rev. D30 (1984) 49.

\item{[18]}
S. D. Ellis, Z. Kunszt and D. E. Soper,
Phys. Rev. Lett.  62 (1989) 726; Phys. Rev. D40, (1989) 2188;
Phys. Rev. Lett.  69 (1992) 1496; \hfil\break\noindent
Z. Kunszt and D. E. Soper, Phys. Rev. D46 (1992) 192.

\item{[19]}
K. J. Eskola and X.-N. Wang
``High $p_T$ Jet Production in $pp$ Collisions'', Preprint HU-TFT-94-44, in
{\it Hard Processes in Hadronic Interactions,} H. Satz and  X.-N. Wang (eds.).

\item{[20]}
A. Donnachie and P. V. Landshoff, Phys. Lett. B332 (1994) 433.
%%%%%``Minijet Production at Small $x$'', Preprint DAMTP 94/28, M/C-TH 94/05.

\item{[21]}
J. D. Bjorken, Phys. Rev.  D27 (1983) 140.

\item{[22]}
K. J. Eskola and X.-N. Wang, Phys. Rev. D49 (1994) 1284.

\item{[23]}
UA2 Collaboration, J. A. Appel {\it et al.}, Phys. Lett.  B 160 (1985)
349;\hfil\break\noindent
UA1 Collaboration, G. Arnison {\it et al.}, Phys. Lett.  B 172 (1986)
461;\hfil\break\noindent
UA1 Collaboration, C. Albajar {\it et al.}, Nucl. Phys. B309 (1988)
405;\hfil\break\noindent
CDF Collaboration, F. Abe {\it et al.},Phys. Rev. Lett.  62 (1989) 613.

\item{[24]}
see Li Xiong's talk in this conference.

\item{[25]}
S. J. Parke and T.R. Taylor; Phys. Rev. Lett. 56 (1986)
2459;\hfil\break\noindent
Z. Kunszt and W. J. Stirling; Phys. Rev. D37 (1988) 2439.

\item{[26]}
V.N. Gribov and L.N. Lipatov, Sov. J. Nucl. Phys. 15 (1972) 438;
\hfil\break\noindent
G. Altarelli and G. Parisi, Nucl. Phys. B126 (1977) 298;\hfil\break\noindent
Yu. L. Dokshitzer, Sov. Phys. JETP 46 (1977) 641.

\item{[27]}
L. V. Gribov, E. M. Levin and M. G. Ryskin,
Phys. Rep.  100 (1983) 1;
Nucl. Phys.  B188 (1981) 555;
Zh. Eksp. Teor. Fiz. 80 (1981) 2132;  \hfil\break\noindent
A. H. Mueller and J. Qiu, Nucl. Phys. B268 (1986) 427.

\item{[28]}
J. Collins and J. Kwieci{\'n}ski, Nucl. Phys. B335 (1990) 89.

\item{[29]}
J. Kwieci{\'n}ski, A. D. Martin, W. J. Stirling and
R. G. Roberts, Phys. Rev.  D42 (1990) 3645.

\item{[30]}
K.J. Eskola, J. Qiu and X.-N. Wang, Phys. Rev. Lett. 72 (1994) 36.

\item{[31]} M. Arneodo, Phys. Rep. 240 (1994) 301.

\item{[32]}
see Raju Venugopalan's talk in this conference.

\item{[33]}
EM Collaboration, M. Arneodo {\it et al.}, Nucl. Phys. B333 (1990) 1;
\hfil\break\noindent
NM Collaboration, P. Amaudruz {\it et al}, Z. Phys. C51 (1991)
387;\hfil\break\noindent
E665 Collaboration, M. R. Adams {\it et al.}, Phys. Rev. Lett. 68 (1992) 3266.

\item{[34]}
K.J. Eskola, Nucl. Phys. B400 (1993) 240.

\item{[35]}
K.J.Eskola and X.-N. Wang, work in progress.

\item{[36]} A Letter of Intent for A Large Ion Collider Experiment (ALICE),
CERN/LHCC/93-16.

\vfill

\eject
\nopagenumbers
{~}
\vfill

\end